\documentclass[useAMS,usenatbib]{mn2e}
\usepackage{graphicx}
\usepackage{amssymb}

\newcommand{\sn}{$\rm S/N$}

\newcommand{\ls}{$\log \sigma_0$}
\newcommand{\sig}{$\sigma_0$}

\newcommand{\simlt}{\lower.5ex\hbox{$\; \buildrel < \over \sim \;$}}
\newcommand{\simgt}{\lower.5ex\hbox{$\; \buildrel > \over \sim \;$}}
\newcommand{\cahk}{$\rm CaHK$}
\newcommand{\mgf}{$\rm Mg4780$}
\newcommand{\nad}{$\rm NaD$}
\newcommand{\tioi}{$\rm TiO1$}

\newcommand{\tioii}{$\rm TiO2$}

\newcommand{\naii}{$\rm Na8190_{SDSS}$}
\newcommand{\cat}{$\rm CaT$}
\newcommand{\cai}{$\rm Ca1$}
\newcommand{\caii}{$\rm Ca2$}
\newcommand{\cnii}{$\rm CN2$}

\newcommand{\cfs}{$\rm C4668$}
\newcommand{\mgi}{$\rm Mg1$}
\newcommand{\mgii}{$\rm Mg2$}
\newcommand{\mgfep}{$\rm [MgFe]'$}
\newcommand{\mgfe}{$\rm [Mg/Fe]$}
\newcommand{\cafe}{$\rm [Ca/Fe]$}
\newcommand{\nafe}{$\rm [Na/Fe]$}
\newcommand{\tife}{$\rm [Ti/Fe]$}
\newcommand{\asfe}{$\rm [O/Fe]$}
\newcommand{\cfe}{$\rm [C/Fe]$}
\newcommand{\nfe}{$\rm [N/Fe]$}
\newcommand{\sife}{$\rm [Si/Fe]$}
\newcommand{\fet}{$\rm Fe3$}
\newcommand{\feff}{$\rm Fe4531$}
\newcommand{\hbo}{$\rm H\beta_o$}

\newcommand{\mgb}{$\rm Mgb5177$}

\newcommand{\caf}{$\rm Ca4227$}

\newcommand{\kms}{\,km\,s$^{-1}$}
\newcommand{\afe}{$\rm [ Mg/Fe]$}
\newcommand{\afep}{$\rm [Z_{Mg}/Z_{Fe}]$}

\newcommand{\gammab}{$\rm \Gamma_b$}
\newcommand{\xfe}{$\rm [X/Fe]$}
\newcommand{\zh}{$\rm [Z/H]$}
\newcommand{\Dix}{$\rm \Delta_{i,X}$}

\title[No correlation of IMF with \afe]
{The Initial Mass Function of Early-type Galaxies: no correlation with \afe}
\author[F.  La   Barbera,  I.  Ferreras,  A.  Vazdekis]
{Francesco  La Barbera$^{1}$\thanks{E-mail:  labarber@na.astro.it},
Ignacio  Ferreras$^{2}$, Alexandre Vazdekis$^{3,4}$\\  
$^{1}$INAF-Osservatorio Astronomico di Capodimonte, sal. Moiariello
16, Napoli, 80131, Italy\\
$^{2}$Mullard Space Science Laboratory, University College London, Holmbury St Mary,
  Dorking, Surrey RH5 6NT, UK\\
$^{3}$Instituto de Astrof\'\i sica de Canarias, Calle V\'\i a L\'actea s/n, E-38205
  La Laguna, Tenerife, Spain\\
$^{4}$Departamento de Astrof\'\i sica, Universidad de La Laguna (ULL), E-38206  La Laguna, Tenerife, Spain
}
\begin{document}

\date{MNRAS Letters, Revised version \today}

\pagerange{\pageref{firstpage}--\pageref{lastpage}} \pubyear{2015}

\maketitle

\label{firstpage}

\begin{abstract}
The Initial Mass Function (IMF) of early-type galaxies (ETGs) has been
found to feature systematic variations by both dynamical and
spectroscopic studies.  In particular, spectral line strengths, based
on gravity-sensitive features, suggest an excess of low-mass stars in
massive ETGs, i.e. a bottom-heavy IMF.  The physical drivers of IMF
variations are currently unknown. The abundance ratio of $\alpha$
elements, such as \afe , has been suggested as a possible driver of
the IMF changes, although dynamical constraints do not support this claim. In
this letter, we take advantage of the large SDSS database.  Our sample
comprises $24,781$ high-quality spectra, covering a large range in
velocity dispersion ($100\!<\!\sigma_0\!<\!320$\,\kms) and abundance
ratio ($-0.1\!<$\afe$<\!+0.4$).  The large volume of data allows us to
stack the spectra at fixed values of \sig\ and \afe .  Our analysis --
based on gravity-sensitive line strengths -- gives a strong
correlation with central velocity dispersion and a negligible
variation with \afe\ at fixed \sig .  This result is robust against
individual elemental abundance variations, and seems not to raise any
apparent inconsistency with the alternative method based on galaxy
dynamics.
\end{abstract}


\section{Introduction}

The stellar IMF, i.e. the mass distribution of stars in a stellar
population {\rm (hereafter SP)} at birth, has been largely assumed to
be universal, mostly because of lack of evidence regarding variations
of the IMF among star clusters and OB associations in our Galaxy
(e.g.~\citealt{Chabrier:2003, Kroupa:2013}).  However, recent studies,
based on independent techniques, such as dynamics
(e.g.~\citealt{Cappellari:2012a, Dutton:2013, Tortora:2013}), lensing
(e.g.~\citealt{Auger:2010}; but see~\citealt{SmithLucey:2013}), and
spectroscopy (e.g.~\citealt{Cenarro:2003, vDC:10, SPI:14}; as well
as~\citealt[hereafter F13 and LB13, respectively]{Ferreras:13,
  LB:13}), have found that the IMF varies systematically with galaxy
mass in ETGs.

The origin of this systematic variation remains
unknown. ~\citet[hereafter CvD12b]{CvD12b} found, for a sample of 38
nearby ETGs, that the IMF normalization (i.e. the stellar M/L,
normalized with respect to a MW-like IMF) correlates more strongly
with the abundance of $\alpha$-elements, such as \afe , than with central velocity
dispersion (\sig), claiming that \afe , which is a proxy for the
star-formation time-scale~\citep{Thomas:05, delaRosa:2011}, might be
the main driver of the IMF in ETGs. However, this finding is
challenged by the fact that dynamical constraints, for a
subsample of 34 ETGs in common between CvD12b and
ATLAS$^{3D}$~\citep{Cappellari:2013}, correlate better with \sig\ 
than \afe~\citep{Smith:2014}. Furthermore, no correlation of dynamical
constraints with { SP} parameters has been detected in 
the $\rm ATLAS^{3D}$ sample \citep{McDermid:2014}.  
Given the importance of constructing a comprehensive theory of a
non-universal { IMF (e.g.}~\citealt{Hopkins:2013, Chabrier:2014,
  Ferreras:2015}), the above inconsistencies call for further
investigations, based on larger samples, of the possible drivers of
IMF variations. In LB13, we assembled a battery of 126 stacked
spectra, with varying \sig\ and \afe , finding that gravity-sensitive
features in ETGs depend very mildly on \afe\, at fixed \sig. Such a
result suggests no correlation between IMF and \afe . In the present
letter, we complement the results of LB13, performing a {\it
  quantitative} analysis of how the IMF slope depends on both \sig\ and
\afe .

\section{Data}
\label{sec:data}
The SPIDER sample~\citep[hereafter LB10]{SpiderI} consists of 39,993
nearby ($0.05 \! < \! z \! < \! 0.095$) ETGs, selected from Data
Release 6 of the Sloan Digital Sky Survey~\citep[SDSS,][]{SDSS:DR6}.
ETGs are defined as bulge-dominated { systems, featuring passive
  spectra within the SDSS fibres}. All SPIDER ETGs have SDSS spectra
available, covering the spectral range from $3800$ to $9200$\,\AA ,
with a spectral resolution $\sigma_{inst} \! \sim \! 60$\,\kms.  For
each spectrum, we use its central velocity dispersion, \sig, from the
SDSS database, and an empirical proxy, \afep , for the
\mgfe\ abundance ratio (see LB13 for details).  The \afep\ is the
difference of total metallicity estimates, \zh , measured from the
\mgb\ and \fet$\rm =(Fe4383+Fe5270+Fe5335)/3$ spectral indices,
respectively.  Total metallicities are measured at fixed age
(estimated with the spectral fitting code
\textsc{STARLIGHT};~\citealt{CID05}), comparing observed line
strengths to predictions of MILES Simple Stellar Population (SSP)
models~\citep{Vazdekis:12}. 
{ As shown in LB13, \afep\ is tightly correlated with 
$\rm [\alpha/Fe]$, derived from SP models taking abundance ratios into
account~\citep{TMJ11}, with $\rm [\alpha/Fe] \sim 0.55$~\afep .  Since
Mg is the reference $\alpha$-element of an SP model, and both \mgb\ and
\fet\ are mostly sensitive to the abundance of Mg and Fe, respectively (see,
e.g.,~\citealt[hereafter JTM12]{JTM12}), in this letter we use
\afe\ rather than ${\rm [\alpha/Fe]}$, with the conversion 
\afe$\sim\!0.55$\afep. Notice that since \afep\ actually measures \mgfe , our
  conclusions do not rely on the assumption that $\alpha$ elements
  should track each other (see, e.g., ~\citealt{KUNT:2010}).  }
Following  LB13   and~F13,  we  restrict  the
analysis to a subsample of 24,781 SPIDER ETGs, { with (i)
$100 \le  \sigma_0 \le
320$\,\kms, } (ii) low internal extinction (i.e.  a color excess estimate
$\rm E(B-V)<0.1$), and (iii) better  signal-to-noise (\sn ) ratio (see
below). We classify the  spectra into 18 narrow bins of  \sig (see LB13 for
details), each bin with  a width of 10\,\kms , except  for the last two, 
for which we adopt  the range  $[260,280]$, and  $[280,320]$\,\kms ,
respectively. For each bin, we only select spectra whose \sn\ ratio is
higher than the lowest quartile of the \sn\ distribution in that bin.

In the present letter, we analyze the median-stacked spectra assembled
by LB13 for this sample, i.e. (i) the 18 ``\sig\ stacks'', obtained by
selecting all ETGs in each \sig\ bin, and (ii) the
``\afep\ sub-stacks'', assembled by selecting, for each bin in \sig,
those ETGs (1) below the 10-th, (2) below the 25-th, (3) between the
25-th and 50-th, (4) between the 50-th and 75-th, (5) above the 75-th,
and (6) above the 90-th percentiles of the \afep\ distribution. This
procedure yields 18 \sig\ stacks, plus $18 \times 6=108$
\afep\ sub-stacks, for a total of $126$ stacked spectra.  At fixed
\sig , the \afep\ stacks span a range of $\delta($\afep$)\sim 0.6$
(corresponding to $\delta($\afe$) \sim 0.3$), allowing us to single out
the effect of \sig\ and abundance ratio on IMF slope.




\section{Analysis}
\label{sec:analysis}

We rely on the extended MILES (MIUSCAT) { SP}
models~\citep{Vazdekis:12} with age (total metallicity) $4\!<\!t\!<\!14$\,Gyr
($-0.4<$\zh$<+0.22$), {
and ten choices of a bimodal IMF~\citep{Vazdekis:1996}, i.e. a
power-law, with slope $\Gamma_b$ in log-mass units, smoothly tapered
to a constant value towards low-mass stars ($\lesssim{0.6M_\odot}$),
with $ 0.3 \le \Gamma_b\le 3.3$.  } Note that \gammab $\sim 1.3$ 
{ gives a good representation of} a Kroupa (MW-like) distribution, while higher values of
\gammab\ correspond to bottom-heavier distributions. For each stacked
spectrum, we derive \gammab\ by a similar procedure to LB13 (see also
F13 and~\citealt[hereafter MN15]{MN:2015}). We minimize the
expression:
\begin{equation}
\chi^2(t, \zh, \Gamma_\mathrm{b}, [X/{\rm Fe}])=
  \sum_i \left[
  \frac{ E_i - E_{M,i} - \sum_X \Delta_{\rm i, X} \cdot [X/{\rm Fe}] }{\sigma_{i}} 
\right]^2,
\label{eq:method}
\end{equation} 
where $E_i$ and $E_{M,i}$ are observed and model line-strengths for a
selected set of spectral features, $[X/{\rm Fe}]$ is the abundance ratio of
different chemical elements (see below), \Dix\ is the sensitivity of
the i-th line strength to a given elemental abundance,
i.e. \Dix$=\delta(E_{M,i})/\delta([X/{\rm Fe}])$, and $\sigma_i$ is the
uncertainty on $E_i$, obtained by adding in quadrature the statistical
error on $E_i$ with its {\it intrinsic} uncertainty (see sec.~6 of
LB13). The $\rm E_{M,i}$ are computed with MIUSCAT models, while the
\Dix\ factors use the publicly available \citet[hereafter CvD12a]{CvD12a} stellar
population models, having solar metallicity, old age ($t\!=\!13.5$\,Gyr),
and a Kroupa IMF. 
The free fitting parameters in Eq.~\ref{eq:method}
are age, metallicity, IMF slope, and elemental abundances.
Both $\rm
E_{M,i}$ and \Dix\ are computed by matching the spectral resolution
of the models to that of each stacked spectrum. Errors on best-fitting
parameters are computed with a bootstrap approach, shifting observed
line-strengths according to their uncertainties. 
{ Notice that by varying the $\Gamma_b$, we actually alter the relative
fraction of low- to high-mass stars in the IMF by changing the
behaviour at higher masses ($\gtrsim{0.6M_\odot}$).  Although this is
different from changes in a unimodal (single power-law) distribution
(e.g. CvD12a), in practice, gravity-sensitive features are only sensitive to the mass
fraction of low-mass stars in the IMF (see figure~19 of LB13).}


\begin{figure*}
\begin{center}
\leavevmode
\includegraphics[width=9.5cm]{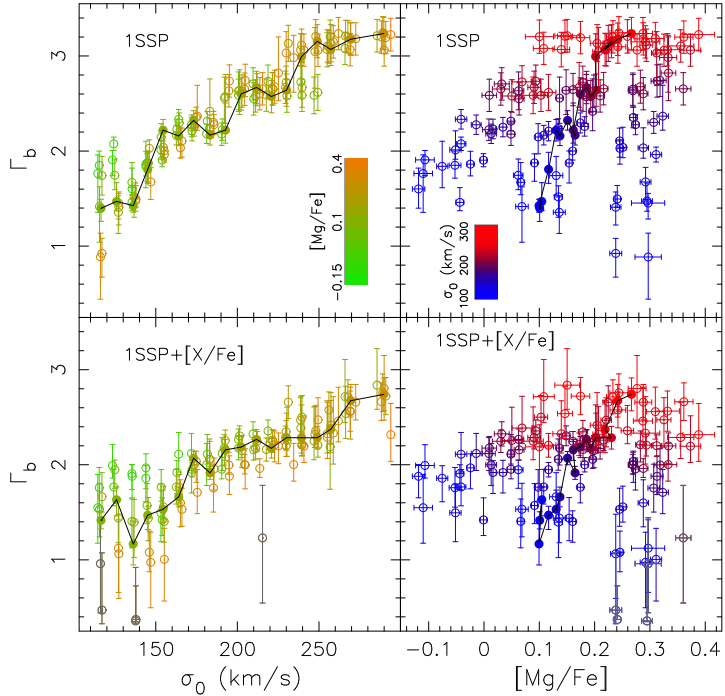}
\end{center}
\caption{ Trends of \gammab\ with \sig\ (left-) and \afe
  (right-panels) for our 126 SDSS-based stacked spectra. Upper and
  lower panels correspond to the 1SSP and 1SSP+\xfe\ fitting methods
  to infer the IMF slope (see Sec.~\ref{sec:analysis}). The
  \sig\ stacks (see Sec.~\ref{sec:data}) are plotted with filled
  circles, and connected by a black curve, while the sub-stacks with
  varying \afe\ are plotted with empty circles. Error bars are quoted
  at the 1\,$\sigma$ confidence level. Blue-through-red colors ({\sl
    right}) encode, as in LB13, the increase of \sig; whereas
  green-through-orange ({\sl left}) correspond to the range in \afe.
  Stacks with a large uncertainty ($>50\%$) on \gammab\ are plotted in
  grey (bottom-panels; see the text).  Notice the tight (poor)
  correlation of \gammab\ with \sig\ (\afe) for both fitting methods.
}
\label{fig:gammab_sig_afe}
\end{figure*}

To test how abundance ratios may affect the estimate of \gammab , we
{ adopt } two alternative methods, differing for the selected set
of spectral features and elemental abundances included in the fitting
(see Tab.~\ref{tab:methods}):
\begin{description}
 \item[{\it 1SSP (Single Burst) method} - ] We use the same set of spectral
   features as in LB13, except for the Calcium triplet, \cat, which
   cannot be measured over the SDSS spectral range for most SPIDER
   ETGs, and is replaced by the \caii\ feature alone (see MN15).  
   { No   \xfe\ term is included in the fits. Indeed, we found that, overall, 
   the leading abundance ratios of gravity-sensitive features have a negligible impact on relative 
   variations of \gammab\ (see, e.g., green and cyan curves in figure~12 of LB13).}
   In contrast to our previous works, we do not apply any \afe\ empirical
   correction to bring observed line-strengths to solar-scale, as (i)
   { the variation of most gravity sensitive-features with $\sigma_0$ is much stronger than
   that with \afe\ (LB13)}, and (ii) our aim, in the present letter, is to
   constrain the IMF as a function of both \sig\ and \afe , hence we
   do not want to apply, {\it a priory}, any \afe\ correction.
 \item[{\it 1SSP+\xfe\ method} - ] We fit all individual abundance
   ratios, whose \Dix\ can be computed with CvD12a models, and adopt a
   much wider set of Lick-based features than in the 1SSP case, making
   sure to include, for each \xfe , at least one index with prominent
   sensitivity to it (based on figure~1 of~JTM12), as
   summarized in Tab.~\ref{tab:methods}. 
   { Notice that the values of \Dix\ may not be constant as a function
  of age (as assumed in Eq.~\ref{eq:method}), as index responses to
  abundance variations could become milder towards younger ages with
  respect to those found in massive ETGs~\citep{Sansom:2013}, implying
  that 1SSP models, instead of 1SSP+\xfe , might describe better our
  low-$\sigma$ stacks (leaving unchanged the conclusions, see below).
  { Moreover, since CvD12a models (i.e. the \Dix\ factors) work at
    fixed Fe, rather than total metallicity (MILES), the \zh\ from
    Eq.~\ref{eq:method} does not necessarily measure the total
    metallicity. This is nevertheless unimportant for the present letter,
    as we study trends with \afep , rather than \zh\ (see Mart\'in-Navarro et al., in
preparation) .}}
\end{description}
We notice that for each stack, the \hbo\ age indicator is
corrected for nebular emission, as in LB13.

\begin{table*}
\centering
\small
 \caption{Methods to derive the IMF slope, \gammab . Col.~1 labels the
   method, while cols.~2 and~3 report the selected list of fitted
   spectral features and elemental abundances (\xfe), respectively.
   For the 1SSP+\xfe\ approach, we include the 1SSP features (from
   LB13), plus other Lick-based abundance-sensitive indices (marked
   with superscripts in col.~2). For each \xfe\ in col.~3, we report
   the superscripts corresponding to the spectral feature(s) with
   prominent sensitivity to it.  }
  \begin{tabular}{c|c|c|c|c}
   \hline
$\rm Method$ & EW & \xfe \\ 
 (1) & (2) & (3) \\
   \hline
1SSP & \hbo, \mgfep, \tioi, \tioii, \mgf, \naii, \caii, \cahk, \nad & none  \\
\\
1SSP+[X/Fe] & \hbo, \mgfep, \tioi, \tioii, \mgf, \naii,  \caii, \cai, \caf$^1$ ,  & 
\cafe$^{1}$, \nafe$^2$, \tife$^3$, \\
            &  \nad$^2$ , \feff$^3$, \mgi$^4$,   \mgii$^5$, \cfs$^6$, \cnii$^7$, \mgb$^8$ &  
            \asfe$^{4, 7}$, \cfe$^{6, 4}$  \\
& & \nfe$^{7}$, \mgfe$^{5, 8}$, \sife$^{4, 5}$ \\
            \hline
  \end{tabular}
\label{tab:methods}
\end{table*}

\section{Results}

Fig.~\ref{fig:gammab_sig_afe} shows the best-fitting \gammab, for all
stacked spectra, { vs.} \sig\ (left-) and
\afe\ (right-panels), respectively. The \sig\ (\afe ) stacks (see
Sec.\ref{sec:data}) are plotted with filled (empty)
circles. Regardless of the fitting method (upper vs. lower panels),
and in particular even when taking abundance ratios into account
(1SSP+\xfe\ method), we find a tight correlation between \gammab\ and
\sig , consistent with previous spectroscopic
studies~(e.g.~CvD12b, LB13). Notice that the quality of the
fits is generally good for all stacks, with a
reduced $\chi^2\!\sim\!2$ ($\sim\!1$) for the 1SSP (1SSP+\xfe )
methods (see sec.~7 of LB13), and -- most importantly for the present
work -- no significant $\chi^2$ dependence on either \sig\ or \afe\ is found. 
The \gammab\--\sig\ relation is shallower for
the 1SSP+\xfe\  case (see below), consistent with
the fact that different \gammab\--\sig\ slopes are found when
adopting different sets of spectral features (F13, \citealt{SPI:14}),
and/or different methodologies (LB13,~\citealt{SPI:15}).  { For} the present work, we do not aim to
establish which method provides a more reliable
\gammab\--\sig\ relation, but, rather, to test the robustness of our
results against two rather extreme cases, namely, the 1SSP method,
where the impact of { abundance} ratios is neglected; 
and the 1SSP+\xfe\ approach, where we attempt to model the effect of
individual elemental abundances, with theoretical (CvD12a) stellar
population models (whose ingredients, such as the stellar atmosphere
calculations, are nevertheless affected by { their own uncertainties}).

Fig.~\ref{fig:gammab_sig_afe} shows that, regardless of the methodology,
the \gammab\ exhibits a far more dispersed correlation with \afe\ than
\sig , i.e.  \afe\ is not the main driver of the IMF variations in ETGs.
The fact that \gammab\ tends to marginally increase with \afe\ mostly
stems from the lack of galaxies with low-\afe\ ($<0.1$)  at the highest \sig\ (see
red symbols in the right panels of Fig.~\ref{fig:gammab_sig_afe}) and
the fact that \afe\ increases with \sig\ in ETGs
(e.g.~\citealt{Thomas:05}). To illustrate this point, we fit a
bivariate relation to the data,
\begin{equation}
\rm 
\Gamma_b = A_1 \log $\sig$ + \rm A_2 $\afe$ + \rm A_3.
\label{eq:reg}
\end{equation}
This approach is the same as \citet[][hereafter S14]{Smith:2014},
with the noticeable difference that in the present work we study the
IMF slope, \gammab, rather than its overall normalization.
Eq.~\ref{eq:reg} allows us to single out, quantitatively, the
contribution of \afe\ and \sig\ to the systematic variations in the
IMF.  We derive $\rm A_1$, $\rm A_2$, and $\rm A_3$ (see
Tab.~\ref{tab:slopes}) with a robust least-square fitting procedure,
minimizing absolute residuals to \gammab , excluding those stacks with
large ($>50\%$) uncertainty on \gammab\ (see grey symbols in
Fig.~\ref{fig:gammab_sig_afe}).  Fig.~\ref{fig:slopes} plots the
best-fitting slopes, $A_1$ and $A_2$, for different fitting methods
together with their bootstrapped confidence contours, with red and
blue colours corresponding to 1SSP and 1SSP+\xfe\ results,
respectively. The green dot and ellipses represent the
1SSP+\xfe\ slopes obtained by replacing \afe\ in Eq.~\ref{eq:reg}
with \mgfe$^*$, i.e. the best-fitting \mgfe\ from 1SSP+\xfe\ fits.  In the 1SSP
method, no significant correlation with \afe\ is found at fixed \sig,
as A$_2$ is consistent with zero at less than 1\,$\sigma$, whereas in
the 1SSP+\xfe\ case, A$_2$ is significantly negative, i.e.
\gammab\ tends to {\sl decrease} with \afe\ at fixed \sig .  Notice
that between these two methods, there is a $\sim 20$\%
difference in $A_1$, with shallower slopes for the 1SSP+\xfe\ case,
consistent with Fig.~\ref{fig:gammab_sig_afe}.
Furthermore, given the range of \ls\ ($\sim 1.2$~dex) and \afe\ ($\sim
0.3$~dex) probed by the data, the values of $\rm A_1$ and $\rm A_2$
for the 1SSP+\xfe\ method imply a narrower variation in the IMF slope
at fixed velocity dispersion, with respect to the variations
associated to \sig. Hence, both approaches agree on \afe\ having a
minor role in the IMF trend of ETGs.

\begin{table}
\centering
\small
 \caption{Best-fitting coefficients, $\rm A_1$, $\rm A_2$, and $\rm
   A_3$ (cols.~2--4), of Eq.~\ref{eq:reg}, for different methods to
   derive \gammab\ (col.~1). The row labeled ``$\cdots$+[Mg/Fe]$^*$'' corresponds to results
   for 1SSP+\xfe\ fits when replacing \afe\ in Eq.~\ref{eq:reg}, with
   \mgfe$^*$ (see the text).}
  \begin{tabular}{c|c|c|c|c}
   \hline
$\rm Method$ & $\rm A_1$ & $\rm A_2$ & $\rm A_3$ \\ 
 (1) & (2) & (3) & (4) \\
   \hline
1SSP            & $3.76 \pm 0.16$ & $-0.08 \pm 0.40$ & $1.95 \pm 0.02$ \\
1SSP+[X/Fe]     & $2.92 \pm 0.24$ & $-0.93 \pm 0.15$ & $1.60 \pm 0.02$ \\
$\cdots$ +[Mg/Fe]$^*$    & $2.80 \pm 0.22$ & $-0.90 \pm 0.14$ & $1.56 \pm 0.02$ \\
   \hline
  \end{tabular}
\label{tab:slopes}
\end{table}

\section{Summary and Discussion}

\begin{figure}
\begin{center}
\leavevmode
\includegraphics[width=7cm]{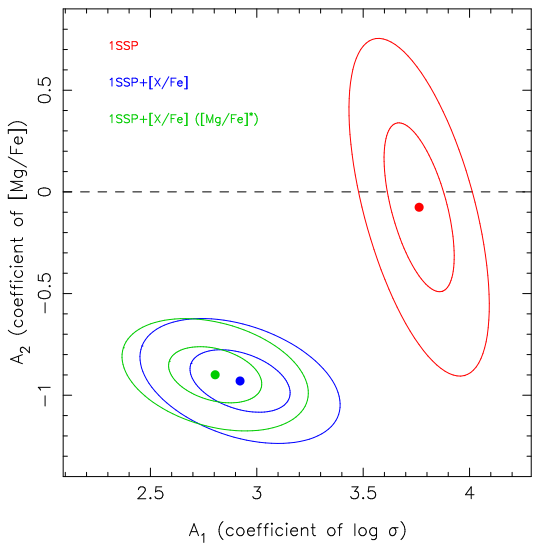}
\end{center}
\caption{ Best-fitting slopes of the bivariate correlation of
  \gammab\ vs. \ls\ and \afe\ (Eq.~\ref{eq:reg}).  Different
  methods/samples are plotted with different colors, as labeled in the
  upper--right of the figure. Ellipses mark 1\,$\sigma$ and
  2\,$\sigma$ normal confidence contours on the slope values. The
  horizontal dashed line represents a zero correlation with \afe\ at
  fixed \sig.  Notice that the \ls\ slope is highly significant in all
  cases, while the \afe\ slope depends on the adopted methodology,
  being fully consistent with zero when no abundance ratios are fitted
  to spectral indices (red dot and ellipses), or mildly {\sl
    anticorrelated} if individual abundances are considered.}
\label{fig:slopes}
\end{figure}

We have assessed in this letter the relative role of central velocity
dispersion and \afe\ abundance ratio as drivers of the systematic
variation found in the IMF of ETGs. We assemble a set of 126
high-quality spectra from a sample of 24,781 nearby ETGs (LB13).  We
split the range in central velocity dispersion
($100\!<\!\sigma_0\!<\!320$\,\kms) into 18 bins. For each bin, i.e.  at
fixed \sig, we create six stacks according to \afe, covering a range
$\Delta$\afe$\sim 0.3$\,dex. For each of the 126 stacks
(i.e. including the original 18 stacks in LB13 only binned with
respect to \sig), we derive the slope of the (bimodal) IMF, \gammab ,
assuming a single burst, and considering both a direct fit to the
line strengths (1SSP method, as in LB13), and an alternative
approach (1SSP+\xfe ), where IMF and individual abundance ratios are
fitted simultaneously.
\begin{description}
 \item[-] We find a very mild variation of \gammab\ with \afe\ at
   fixed \sig , implying that \afe\ plays a subdominant role in the
   variation of the IMF slope among ETGs.
 \item[-] On the contrary, we find a strong, highly significant
   ($>\!10\,\sigma$), correlation of the IMF slope with \sig ,
   regardless of the adopted methodology.
\end{description}
Since  for  a   fixed  IMF  functional  form  (such   as  the  bimodal
distribution), \gammab\  is proportional to the  overall normalization
(i.e. the  mass-to-light ratio  normalized to  that for  a Kroupa-like
distribution), our  results contrast with  the claim of  CvD12b, that
the IMF should  correlate more strongly with \afe\ than  \sig. We note
that the models  of CvD12b use [Fe/H] and not  [Z/H] as an ``overall''
metallicity  parameter. Therefore  their [Mg/Fe]  variations could  be
related to total  metallicity. S14 found that for a  sample of 34 ETGs
in  common   between  CvD12b  and  Atlas$^{\rm   3D}$,  the  dynamical
constraints to  the IMF appeared  not to  be consistent with  the line
strength analysis of CvD12b.  Dynamical models correlate with velocity
dispersion (\sig ) and not with \afe . Therefore, our results -- based
on  spectral indices  -- are  qualitatively more  consistent with  the
results  from  dynamical  modeling.  In  particular,  the  fact  that
\gammab\ decreases slightly with \afe , at fixed \sig , when using the
1SSP+\xfe\ fitting method, is similar  to the results in S14 regarding
the   dynamical   constraints   (see    fig.~4   of   S14,   and   our
Fig.~\ref{fig:slopes}).   At  the  moment,  the origin  of  the  above
discrepancies  is  not  clear.  As already  noticed  by  S14,  several
possible effects might contribute, including (i) sample size/selection
issues, as  we analyze a sample  of 126 stacked spectra,  while CvD12b
studied 38 (individual) galaxies; (ii) methodology, as we rely on line
strengths, rather  than spectral fitting, and  adopt (MIUSCAT) stellar
population models with varying total  metallicity (in contrast to CvD12a
models); (iii)  systematics, on either stellar  population models (for
spectroscopic  constraints)  or  the modeling  of  the  (dark-)matter
distribution  in   ETGs  (for  dynamical  constraints);   (iv)  radial
variations of the  IMF, as the SDSS spectra cover  $\sim \rm 1R_{eff}$
spatial scale,  in contrast  to the R$_{\rm  eff}/8$ scale  of CvD12b.
The  latter aspect  is particularly  relevant in  light of  our recent
finding  (MN15)  that the  IMF  slope  decreases with  galacto-centric
distance   in  massive   ETGs  (see   also~\citealt{Pastorello:2014}),
implying that local and global constraints  to the IMF might result in
different trends. We also remark that  the present work does not imply
that the central  \sig\ (an integrated measurement)  is the ``causal''
driver of the  IMF in ETGs, but, rather, that  \afe\ (possibly related
to  the timescale  of star  formation) is  not driving  the systematic
variations of  the IMF, at  least over  a $\sim \rm  1R_{eff}$ spatial
scale. In particular, given  the tight correlation between metallicity
and \sig\  in ETGs, and the  narrow dynamical range of  metallicity in
our stacked spectra at fixed \sig\  (LB13) , the values of \gammab\ in
Fig.~\ref{fig:gammab_sig_afe} can also be considered to correlate with
(total) metallicity.  Hence, our integrated  spectra cannot be used to
infer  the   physical  driver   of  the  trend.    Spatially  resolved
spectroscopic studies, extended to large samples of galaxies, with the
aid of further advances in the modeling of stellar populations, should
help in the future to address  this issue (Mart\'in-Navarro et al., in
preparation).

\section*{Acknowledgments}
{ We thank Dr. M.~Beasley, and the anonymous referee, for
  interesting comments and suggestions. We acknowledge the use of
  SDSS data (http://www.sdss.org/collaboration/credits.html), and
  support from grant AYA2013-48226-C3-1-P from the Spanish Ministry of
  Economy and Competitiveness (MINECO).}

\label{lastpage}

\end{document}